\documentclass[11pt,a4paper]{article}
 
\usepackage{amsmath,amsthm,amssymb}

\usepackage{cite}

\pdfoutput=1

\usepackage{graphics,graphicx}
\usepackage{epsfig}
\usepackage{multicol}
\usepackage{color}
\makeatletter
\@addtoreset{equation}{section}
\makeatother

\usepackage{braket}
\usepackage{bm}

\begin{document} 
\sloppy

\begin{center}
\LARGE{{\bf An Introduction to Noncommutative Physics}}
\end{center}

\begin{center}
\large{Shi-Don Liang${}^{a,b}$\footnote{stslsd@mail.sysu.edu.cn} and Matthew J. Lake${}^{c,,d,a,e,f*}$\footnote{matthewjlake@narit.or.th}}
\end{center}
\begin{center}
\emph{$^{a}$School of Physics, Sun Yat-Sen University, \\ Guangzhou 510275, People’s Republic of China \\}
\emph{$^{b}$ State Key Laboratory of Optoelectronic Material and Technology, \\ Guangdong Province Key Laboratory of Display Material and Technology, \\ Sun Yat-Sen University, Guangzhou 510275, People’s Republic of China\\} 
\emph{$^{c}$National Astronomical Research Institute of Thailand, \\ 260 Moo 4, T. Donkaew,  A. Maerim, Chiang Mai 50180, Thailand \\}
\emph{$^{d}$Department of Physics and Materials Science, \\ Faculty of Science, Chiang Mai University, \\ 239 Huaykaew Road, T. Suthep, A. Muang, Chiang Mai 50200, Thailand \\}
\emph{$^{e}$Department of Physics, Babe\c s-Bolyai University, \\ Mihail Kog\u alniceanu Street 1, 400084 Cluj-Napoca, Romania \\}
\emph{$^{f}$Office of Research Administration, Chiang Mai University, \\ 239 Huaykaew Rd, T. Suthep, A. Muang, Chiang Mai 50200, Thailand \\}
\vspace{0.1cm}
\end{center}

\begin{abstract}
In recent years, many new developments in theoretical physics, and in practical applications rely on different techniques of noncommutative algebras.
In this review, we introduce the basic concepts and techniques of noncommutative physics in a range of areas, including classical physics, condensed matter systems, statistical mechanics, and quantum mechanics, and we present some important examples of noncommutative algebras, including the classical Poisson brackets, the Heisenberg algebra, Lie and Clifford algebras, the Dirac algebra, and the Snyder and Nambu algebras.
Potential applications of noncommutative structures in high-energy physics and gravitational theory are also discussed.
In particular, we review the formalism of noncommutative quantum mechanics based on the Seiberg--Witten map and propose a parameterization scheme to associate the noncommutative parameters with the Planck length and the cosmological constant. We show that noncommutativity gives rise to an effective gauge field, in the Schr\"{o}dinger and Pauli equations.
This term breaks translation and rotational symmetries in the noncommutative phase space, generating intrinsic quantum fluctuations of the velocity and acceleration, even for free particles.
This review is intended as an introduction to noncommutative phenomenology for physicists, as well as a basic introduction to the mathematical formalisms underlying these effects.
\end{abstract}


\newpage
\tableofcontents 
\newpage

\section{Introduction} \label{Sec.1}

Different physical theories depend on different mathematical structures.
For example, classical Hamiltonian mechanics is defined on a symplectic manifold, and relativistic gravity theories describe the dynamics of pseudo-Riemann geometries, while quantum theories are defined by using complex vector spaces~\cite{Takhtajan1}.
Noncommutativity naturally arises in various formalisms, tracing back to the descriptions of angular momentum and work in classical mechanics, through to the Heisenberg algebra and its associated uncertainty relations in canonical quantum mechanics, and~on to more speculative recent theories regarding the noncommutative nature of spacetime at the Planck scale~\cite{Douglas,Szabo,Konechny,Rosenbaum}.
The latter include the Snyder algebra~\cite{Snyder}, Heisenberg--Weyl algebra, various types of Lie algebra, and~the so-called noncommutative phase space algebra~\cite{Douglas,Gouba}.

Snyder proposed a five-dimensional noncommutative spacetime, with~global Lorentz invariance, to~remove singularities in particle physics without renormalization techniques~\cite{Snyder}.
His model was generalized to curved spacetimes by C. N. Yang, in~order to include gravitational effects~\cite{Yang}.
These early studies were among the first to suggest that quantized spacetime should be described by a kind of noncommutative geometry (NCG), which, it is hoped, can provide a self-consistent formalism to unify quantum theory with \mbox{gravity~\cite{Carlo,Thomas,Fredenhagen}.}
Thus, it is believed that NCG could play a vital role in removing the infinities and singularities that unavoidably emerge in both particle physics and~cosmology.

{In addition, the~recent discoveries of dark energy and dark matter in cosmological observations lead to many puzzles, and the fundamental physics behind these phenomena is not well understood.}
There are many candidate theories, and, even though dark energy can be interpreted phenomenologically as a cosmological constant, there is still no direct experimental evidence capable of determining the physical mechanism that gives rise to it~\cite{Sabino,Peebles}.
One possibility is that the small but nonzero vacuum energy emerges from the quantum fluctuations of the spacetime background at the Planck scale, which is expected in the framework of noncommutative field theory and quantum gravity models based on noncommutative 
geometry~\cite{Connes,Connes1,Doplicher}.

In less speculative fields, noncommutativity also plays a vital role in describing a range of  important physical phenomena.
{In condensed matter physics, it may be shown that a two-dimensional electronic system, immersed in a strong magnetic field, is equivalent to its free electron counterpart, formulated in a noncommutative phase 
space~\cite{Francois,Kovacik,Bellucci,Gamboa,Gamboa1,Ezawa}. 
Thus, by~generalizing the canonical Heisenberg algebra to include nontrivial space--space and/or momentum--momentum commutation relations, this so-called noncommutative quantum mechanics (NCQM) can successfully model  known phenomena \cite{Galikova:2015eyn}, such as the Aharonov--Bohm effect~\cite{Das,Liang,Liang1,Masud}, the~quantum Hall effect~\cite{Ezawa,Kokado,Basu}, the~existence of magnetic monopoles~\cite{Francois} and the Berry phase~\cite{Bertolami1,Falomir}, by using a different language and mathematical~formalism.}

Currently, there are several schemes used to implement such phase space noncommutativity~\cite{Gouba,Sivasubramanian,Kokado,Vilela,Harko}, including the 
Seiberg--Witten (SW) map~\cite{Seiberg,Seiberg1,Bastos1}, the~Moyal product formalism~\cite{Szabo} and the Wigner--Weyl phase space approach~\cite{Bastos2,Bastos2a,Bastos2b,Ho,Chaichian:2002ew}. 
Based on these noncommutative algebras, the~Heisenberg uncertainty principle can be extended to include new space--space and momentum--momentum uncertainty relations, which are clearly of great phenomenological interest to physicists. 
These physical phenomena also stimulate mathematical interest in noncommutative algebras and noncommutative 
geometry~\cite{Connes,Connes1,Doplicher}.

An example of a new mathematical formalism, with~potential applications in high-energy physics and gravitational theory, is the Nambu generalization of symplectic geometry, in~which the Poisson algebra is generalized to the so-called Nambu bracket, yielding generalized Hamiltonian~\cite{Nambu,Bayen,Estabrook} and Lagrangian mechanics~\cite{Mukunda,Kalnay}.
The quantisation of the classical Nambu brackets generates a deformed Heisenberg algebra~\cite{Curtright,Awata,Sahoo,Zachos}, which leads to new physical phenomenology~\cite{Hirayama1,Hirayama2,Minic}, and~to a new phase space structure, which is also of interest to mathematicians~\cite{Angulo,Takhtajan,Fecko}, and may have applications in string theory \cite{Sheikh-Jabbari:2004fiz,Sheikh-Jabbari:2006nsl}.

In this paper, we present an accessible introduction to noncommutative physics in~a pedagogical format with~a strong focus on phenomenology.
For completeness, and~so that the text can be read as a self-contained reference, we also include brief summaries of the basic mathematical formalisms needed to implement noncommutative structures in~a range of example systems.
{We begin by reviewing important examples of noncommutative phenomena in physics in~Section~\ref{Sec.2}, including the canonical Poisson brackets of classical mechanics, permutation symmetries in statistical mechanics, and~the classical and quantum Hall effects.}

In Section~\ref{Sec.3}, we give an overview of some known noncommutative algebras, including the Poisson algebra in classical symplectic geometry, the~Heisenberg, Lie, Clifford, and~Dirac algebras in canonical quantum theory, the~Snyder and Nambu algebras in theoretical physics, and~the deformed Heisenberg algebra used to generate models of NCQM \cite{Gouba}.

In Section~\ref{Sec.4}, we give a detailed presentation of NCQM, based on the deformed Heisenberg algebra, and~review its phenomenological implications for gravitational theories and high-energy particle physics.
Based on the SW map, we give the Heisenberg representation of the Schr\"{o}dinger, Heisenberg,  and Pauli equations, and~consider some basic properties of the model, including the existence of anomalous velocity and acceleration terms in the free-particle dynamics induced by the noncommutativity of the background.
We propose a parameterization scheme that associates the noncommutative parameters of the space--space and momentum--momentum commutators, respectively, with~the Planck length and the cosmological constant and~discuss its implications for particle physics and~cosmology.

However, in~this pedagogical introduction to noncommutative physics we do not discuss Conne's approach to noncommutative geometry, K theory, noncommutative field theory, the~Moyal star product technique, or~$\kappa-$deformed symmetries of spacetime.
For details of these (more mathematical) approaches, readers are referred to \cite{Connes,Connes1,Doplicher,Emil,Giovanni,Alessandra,Michele,Chaichian:2001py} and references therein. 
Likewise, we do not give a detailed exposition of the role of noncommutative geomtry in string theory. 
Further information on this topic, including the role of NCQM in string phenomenology, can be found in \cite{Seiberg1,Lizzi:1999dd,Blumenhagen:2014sba,Ardalan:1998ce,Chaichian:2001pw}.
We summarise our conclusions, and~offer a few opinions on the outlook for noncommutative geometry in physics research, in~Section~\ref{Sec.5}.

\section{Noncommutative Phenomena in Physics: Important~Examples} \label{Sec.2}

In this Section, we review important examples of noncommutative phenomena in classical and quantum physics.

\subsection{Noncommutativity in Classical~Physics}  \label{Sec.2.1}

The best known example of noncommutativity in classical physics comes from Hamiltonian mechanics, in which the Poisson bracket is defined as
\begin{equation}\label{PSB1}
\left\{f,g\right\}:=\frac{\partial f}{\partial q^i}\frac{\partial g}{\partial p_i}-\frac{\partial f}{\partial p_i}\frac{\partial g}{\partial q^i}.
\end{equation}
This relation implies that the generalised coordinates, $q^i$, $i \in \left\{1,2, \dots d\right\}$, where $d$ is the dimensionality of the system, and their corresponding canonical momenta, 
$p_i = \partial L/\partial q^i$, obey the commutation relations~\cite{Deriglazov} 
\begin{eqnarray} \label{CCR2A}
\left\{q^i , q^j \right\} &=& 0, 
\end{eqnarray}
\begin{eqnarray} \label{CCR2B}
\left\{p_i , p_j \right\} &=& 0, 
\end{eqnarray}
\begin{eqnarray} \label{CCR2C}
\left\{q^i , p_j \right\} &=& \delta^{i}{}_{j},
\end{eqnarray}
where $\delta^{i}{}_{j}$ is the Kronecker delta. 
These relations are called the fundamental Poisson brackets.
Together, the~coordinates and canonical momenta define the phase space of the system.
All physical quantities are expressed in terms of maps on the phase space and the evolution of the system follows the integral curves generated 
by the flow of the Hamiltonian vector field, 
\begin{equation}\label{PB2}
\dot{\gamma}=\{\gamma,\mathcal{H}\}.
\end{equation}
The physical quantity $\gamma$ is conserved if $\{\gamma,\mathcal{H}\}=0$ \cite{Deriglazov}.

\subsection{Permutation Symmetries in Statistical~Mechanics}  \label{Sec.2.2}

{In statistical physics, the~statistical behavior of multiparticle systems depends sensitively on the permutation symmetries of elementary particles.}
The mean occupation numbers for free particles in thermal equilibrium states are given by
\begin{equation}
n=e^{-(\varepsilon-\mu)/k_BT}
\end{equation}
for classical particles, and~\begin{equation}
n=\left\{
\begin{array}{cc}
\frac{1}{e^{(\varepsilon-\mu)/k_BT}-1} & \textrm{for bosons,} \\
\frac{1}{e^{(\varepsilon-\mu)/k_BT}+1} & \textrm{for fermions}
\end{array}\right.
\end{equation}
in quantum physics, where $T$ is the temperature of the system, $k_B$ is Boltzmann's constant, and $\mu$ is the chemical potential.
The differing bosonic and fermionic distributions, known as the Bose--Einstein (BE) 
distribution and the Fermi--Dirac (FD) distribution, respectively, are governed by different algebras.
In the particle number representation, the~particle creation and annihilation operators for bosons and fermions obey, 
respectively, 
\begin{eqnarray} \label{AAA-A}
\left[\hat{b}_i,\hat{b}_j^\dagger\right] &=& \delta_{ij}\hat{\mathbb{I}}, \quad \left[\hat{b}_i,\hat{b}_j\right] =\left[\hat{b}_i^\dagger,\hat{b}_j^\dagger\right] = 0 
\, \textrm{(BE)}, 
\end{eqnarray}
\begin{eqnarray} \label{AAA-B}
\left\{\hat{a}_i,\hat{a}_j^\dagger\right\} &=& \delta_{ij} \hat{\mathbb{I}}, \quad \left\{\hat{a}_i,\hat{a}_j\right\}=\left\{\hat{a}_i^\dagger,\hat{a}_j^\dagger\right\} =0 
\, \textrm{(FD)},
\end{eqnarray}
where $\hat{\mathbb{I}}$ is the identity operator, $[A,B] \equiv AB-BA$ denotes the commutator of $AB$ and $B$, and $\left\{A,B\right\}\equiv AB+BA$ is the anticommutator.
The anticommutative property of fermion operators ($\hat{a}_i\hat{a}_j=-\hat{a}_j\hat{a}_i, \hat{a}_i^2=0$) is known as a Grassmann algebra.
The relationship between the Fermi--Dirac distribution, spin, and~the Grassmann algebra, is called the spin-statistics theorem~\cite{Takhtajan}.

\subsection{The Heisenberg~Algebra}  \label{Sec.2.3}

To understand the discrete atomic spectrum of hydrogen, Bohr proposed an orbital model of the atom in which discrete spectra are generated by the transitions of electrons between orbits with different energy levels, $\omega_{nm}=(E_n-E_m)/\hbar$, where $E_{n(m)}$ label the discrete energy levels and $\hbar$ is the reduced Planck's constant. 
Heisenberg proposed an equation of motion, now known as the Heisenberg equation, from~which Bohr's empirical formula could be derived in a more rigorous way,
\begin{equation}\label{HEM1}
i\hbar\frac{d\hat{O}}{dt}=[\hat{O},\hat{H}],
\end{equation}
where $\hat{O}$ is the operator corresponding to any physical observable and $\hat{H}$ is the Hamiltonian operator of the system.
For any system, all observables, with the exception of spin (discussed in Section~\ref{Sec.2.2} above), 
are functions of the canonical position and momentum operators, $\hat{x}^i$ and $\hat{p}_j$.
These are governed by the noncommutative algebra,
\begin{equation}\label{HAGA}
[\hat{x}^i,\hat{p}_j] =i\hbar\delta^{i}{}_{j}\hat{\mathbb{I}}, \ \  [\hat{x}^i,\hat{x}^j] = 0, \ \ [\hat{p}_i,\hat{p}_j] = 0,
\end{equation}
where $i,j \in \left\{1,2,3 \right\}$ and $(x^1,x^2,x^3) \equiv (x,y,z)$ denote global Cartesian coordinates in three-dimensional Euclidean space, $\mathbb{R}^3$.
The noncommutativity of the position and momentum variables leads directly to the Heisenberg uncertainty 
relation,
\begin{equation}\label{HUP}
\Delta x^i\Delta p_j\geq \frac{\hbar}{2}\delta^{i}{}_{j},
\end{equation}
and Equation~(\ref{HAGA}) is known as the Heisenberg algebra. 
The uncertainty relation reveals the intrinsic quantum fluctuations caused by the wave--particle duality of matter in the microscopic world.
This form of noncommutativity not only leads to the emergence of quantum mechanics, but~was also the original inspiration stimulating many attempts to extend noncommutative concepts to different fields in physics~\cite{Douglas,Gouba}.

\subsection{The Classical and Quantum Hall~Effects}  \label{Sec.2.4}

The current density in a conductor is given by
\begin{equation}\label{CDF1}
J_{i}=\sigma_{ij}E^j,
\end{equation}
where $\sigma_{ij}$ are the components of the conductivity matrix and $E^j$ are the components of the applied electric field.
In the equilibrium state, $\sigma_{xx}=\sigma_{yy}=0$,
the transverse {conductivity} obeys the noncommutative relation, 
\begin{equation}\label{NCHC}
\sigma_{xy} -\sigma_{yx}=2\sigma_H,
\end{equation}
where $\sigma_H=n_{e}e/B$, $e$ is the charge of the electron, $n_e$ is the density of electrons per unit area, and $B = |{\bf B}|$ is the magnitude of the magnetic field strength~\cite{Ezawa}. 
Discovered by Edwin Hall in 1879, this is the canonical example of noncommutativity in classical electrodynamics.
Practically, the~noncommutative relation (\ref{NCHC}) can be applied to detect the types of charge carriers in semiconductors, namely, electrons $(-)$ or holes $(+)$.

In the quantum Hall effect, which occurs in the presence of strong magnetic fields, the quantized Hall conductivity emerges, 
$\sigma_{ij}=\nu e^2\epsilon_{ij}/\hbar$, where $\epsilon_{ij}$ is the antisymmetric (Levi--Civita) tensor. 
Here, $\nu$ is the filling factor defined as  
$\nu= n_e/n_B 
\in \mathbb{N}$, where $n_B=2\pi eB/\hbar$ is the maximum number of electron states per unit area of a single Landau level~\cite{Ezawa}.

In a strong constant magnetic field, the electron feels a Lorentz force and undergoes circular motion in the plane perpendicular to the magnetic field lines.
The planar coordinates, $(x,y)$, may be decomposed into the ``guiding center'' $(X,Y)$ and the ``relative coordinate'' $(R_x,R_y)$ parts, such that $\mathbf{x}=\mathbf{X}+\mathbf{R}$, where $R_x=-P_Y/eB$, $R_y=P_X/eB$ and $(P_X,P_Y)$ denote the covariant momenta.
The guiding center coordinates are then given by~\cite{Ezawa} 
\begin{eqnarray} \label{XY1-A}
\hat{X} &=& \hat{x} - \frac{1}{eB}\hat{P}_Y, 
\end{eqnarray}
\begin{eqnarray} \label{XY1-B}
\hat{Y} &=& \hat{y}+\frac{1}{eB}\hat{P}_X,
\end{eqnarray}
where $[\hat{x},\hat{P}_X]=[\hat{y},\hat{P}_Y]=i\hbar\hat{\mathbb{I}}$.
We then have
\begin{eqnarray} \label{XYPP1-A}
\left[\hat{X},\hat{Y}\right] &=& i \ell_B^2\hat{\mathbb{I}}, 
\end{eqnarray}
\begin{eqnarray} \label{XYPP1-B}
\left[\hat{P}_X,\hat{P}_Y\right] &=& i \frac{\hbar^2}{\ell_B^2} \hat{\mathbb{I}},
\end{eqnarray}
and
\begin{eqnarray} \label{XYPP2-A}
\left[\hat{X},\hat{P}_X\right] &=& \left[\hat{X},\hat{P}_Y\right]=0, 
\end{eqnarray}
\begin{eqnarray} \label{XYPP2-B}
\left[\hat{Y},\hat{P}_X\right] &=& \left[\hat{Y},\hat{P}_Y\right]=0,
\end{eqnarray}
where
\begin{eqnarray} \label{l_B}
\ell_B=\sqrt{\frac{\hbar}{eB}}
\end{eqnarray}
is called the magnetic length, which describes the fundamental scale in the quantum Hall~system.

To solve the quantum Hall system, we introduce two kinds of boson operators~\cite{Ezawa}, 
\begin{eqnarray} \label{aabb1-A}
\hat{a} &=& \frac{\ell_B}{\sqrt{2}\hbar}(\hat{P}_X+i\hat{P}_Y), 
\end{eqnarray}
\begin{eqnarray} \label{aabb1-B}
\hat{a}^\dagger &=& \frac{\ell_B}{\sqrt{2}\hbar}(\hat{P}_X-i\hat{P}_Y),
\end{eqnarray}
\begin{eqnarray} \label{aabb1-C}
\hat{b} &=& \frac{1}{\sqrt{2}\ell_B}(\hat{X}+i\hat{Y}), 
\end{eqnarray}
\begin{eqnarray} \label{aabb1-D}
\hat{b}^\dagger &=& \frac{1}{\sqrt{2}\ell_B}(\hat{X}-i\hat{Y}),
\end{eqnarray}
which obey the commutation relations:
\begin{eqnarray} \label{aabb2-A}
\left[\hat{a},\hat{a}^\dagger\right] &=& \left[\hat{b},\hat{b}^\dagger\right]=1, 
\end{eqnarray}
\begin{eqnarray} \label{aabb2-B}
\left[\hat{a},\hat{b}\right] &=& \left[\hat{a},\hat{b}^\dagger\right]=0, 
\end{eqnarray}
\begin{eqnarray} \label{aabb2-C}
\left[\hat{a}^\dagger,\hat{b}\right] &=& \left[\hat{a}^\dagger,\hat{b}^\dagger\right]=0,
\end{eqnarray}
and the Fock vacuum is defined by the conditions $\hat{a}|0\rangle=0$ and $\hat{b}|0\rangle=0$.
The Fock states are explicitly constructed as
\begin{equation}\label{FS1}
|N,n\rangle=\sqrt{\frac{1}{N!n!}}(\hat{a}^\dagger)^N(\hat{b}^\dagger)^n|0\rangle,
\end{equation}
and the orthonormal completeness relations are
\begin{eqnarray} \label{NnNn-A}
\langle M,m|N,n \rangle &=& \delta_{MN}\delta_{mn}, 
\end{eqnarray}
\begin{eqnarray} \label{NnNn-B}
\sum_{N,n}|N,n\rangle\langle N,n| &=& 1.
\end{eqnarray}

Consequently, the~Hamiltonian can be rewritten as
\begin{equation}\label{Haa}
\hat{H}=\left(\hat{b}^\dagger \hat{b}+\frac{1}{2}\right)\hbar\omega_c,
\end{equation}
where $\omega_c={\hbar}/({m_e\ell_B^2})={eB}/{m_e}$, 
and $m_e$ is the effective mass of the quantum Hall system.
The corresponding eigenenergies (Landau levels) are then obtained as
\begin{equation}\label{EHaa}
E_N=\left(N+\frac{1}{2}\right)\hbar\omega_c.
\end{equation}

Each level is multiply degenerate and the state $|N,n\rangle$, given in Equation~(\ref{FS1}), is called the $n$th Landau site in the $N$th Landau level.
Each Landau site occupies an area $\Delta A=2\pi \ell_B^2$ in real space.
The density of states is given by
\begin{equation}\label{DSS1}
\rho_\phi=\frac{1}{2\pi\ell_B^2}=\frac{B}{\phi_D},
\end{equation}
where $\phi_D= 2\pi\hbar/e$ 
is the Dirac flux quantum.
The quantization of the Hall conductance then leads to the relation~\cite{Ezawa}
\begin{equation}\label{nphi}
\nu=\frac{n_e}{\rho_\phi}=2\pi\ell_B^2n_e.
\end{equation}

In Section~\ref{Sec.4.8} below, we show how constructing a quantum Hall system in a noncommutative phase space modifies the effective magnetic field, giving  
$B^{\rm nc}=B+B^\eta$, where $B^\eta$ is related to the noncommututivty parameter of the momentum--momentum commutation relations, $\eta$.
In the proposed parameterization scheme (see Section~\ref{Sec.4.8}), $\eta = m_{\rm dS}^2c^2 \equiv \hbar^2\Lambda/3$, where $m_{\rm dS} 
= (\hbar/c)\sqrt{\Lambda/3}$ is the de Sitter mass, $\Lambda \simeq 10^{-52} \, {\rm cm^{-2}}$ is the cosmological constant~\cite{Peebles}, and $c$ denotes the speed of light.
The effective magnetic length becomes $\ell_B^{\rm nc}=\sqrt{\frac{\hbar}{e(B+\hbar\Lambda)}}$ and the {quantized Hall conductivity 
is $\sigma_{ij}=\nu e^2\epsilon_{ij}/h$, where $\nu=2\pi(\ell_{B}^{\rm nc}){}^{2}n_e$, and $h$ is Planck's constant, $2\pi \hbar$. 
However, since $B^\eta\ll B$, the~noncommutative effect is too small to be observable with current technology~\cite{Gouba,Liang,Liang1}.

\section{An Overview of Noncommutative~Algebras}  \label{Sec.3}

Noncommutative phenomena are governed by noncommutative algebras. 
Here, we present several typical examples of noncommutative algebras in~physics.

\subsection{The Poisson Algebra in Symplectic~Space}  \label{Sec.3.1}

In classical mechanics, the Poisson bracket corresponds to the quantum mechanical commutator, in~the limit $\hbar\rightarrow 0$ of a canonical quantization scheme.
The Poisson bracket can be regarded as a generalized product that defines an algebra (the Poisson algebra), which gives rise to the symplectic structure of Hamiltonian mechanics in the canonical phase space.
The canonical phase space is defined, formally, as~the cotangent bundle, $T^*Q^n$, of~the configuration space, $Q^n$, with~the symplectic structure given by~\cite{Deriglazov}
\begin{equation}\label{SST1}
\Omega=dq^i\wedge dp_i,
\end{equation}
where $A \wedge B = A \otimes B - B \otimes A$ is the Cartan wedge product.
In the matrix representation of the symplectic structure, the~generalized coordinates and momenta are redefined as $\left\{z_\alpha\right\}=\left\{z_1,\cdots,z_{2n}\right\}:=\left\{q_1,\cdots,q_n,p_1,\cdots,p_n\right\rfloor$, where $i = 1,\cdots,n$ for $q$ and $p$, while $\alpha=1,\cdots,2n$ for $z$.
{The symplectic 2-form $\Omega$ is defined as a bilinear map, $\Omega: Z\times Z\rightarrow \mathbb{R}$, where $Z$ is a linear vector space~\cite{Deriglazov}, i.e.,
\begin{equation}\label{2SS}
\Omega(z,z)=z^T J z,
\end{equation}
where $z$ is a column vector with $2n$ components $z\in Z$, the~subscript $T$ denotes the transpose, and~$J=(J_{\alpha\beta})$ is the $2n\times 2n$ symplectic matrix
\begin{equation}\label{SPM1}
J=\left(
\begin{array}{cc}
0 & I \\
-I & 0
\end{array}
\right).
\end{equation}
Here, $0$ denotes the $n\times n$-zero matrix, and $I$ is the $n\times n$ unit matrix.}
It can be seen that $J$ is skewed and regular, $\det J=1$.

For any pair differentiable functions on the phase space, $f,g\in \mathcal{F}(T^*Q)$, one has 
\begin{eqnarray}\label{SST1}
df\wedge dg = \frac{\partial f}{\partial q^i}\frac{\partial g}{\partial p_i}dq^i\wedge dp_i  
=  \frac{1}{2}\left(\frac{\partial f}{\partial q^i}\frac{\partial g}{\partial p_i}-\frac{\partial f}{\partial p_i}\frac{\partial g}{\partial q^i}\right)dq^i dp_i,
\end{eqnarray}
and the Poisson bracket is defined by Equation~(\ref{PSB1}). 
The Poisson bracket is skew-symmetric and bilinear, and~obeys the Leibniz rule and the {Jacobi identities}, namely~\cite{Deriglazov}
\begin{eqnarray} \label{PSBK2-A}
\left\{f , g \right\} &=& -\left\{g , f \right\}, 
\end{eqnarray}
\begin{eqnarray} \label{PSBK2-B}
\left\{\alpha f + \beta g ,  h\right\} &=& \alpha\left\{f, g  \right\}+\beta\left\{f, h\right\}, 
\end{eqnarray}
\begin{eqnarray} \label{PSBK2-C}
\left\{f , \alpha g +\beta h\right\} &=& \alpha\left\{f, h  \right\}+\beta\left\{g, h\right\}, 
\end{eqnarray}
\begin{eqnarray} \label{PSBK2-D}
\left\{f , gh \right\} &=& \left\{f , g \right\} h+ g\left\{f , h \right\}, 
\end{eqnarray}
\begin{eqnarray} \label{PSBK2-E}
\left\{\left\{f , g \right\},h\right\} &+& \left\{\left\{g , h \right\},f\right\}+\left\{\left\{h , f \right\},g\right\}=0,
\end{eqnarray}
where $\alpha$ and $\beta$ are real~constants.

\subsection{From Poisson to~Heisenberg} \label{Sec.3.2}

The canonical quantization procedure is based on a correspondence between the classical Poisson brackets and the Heisenberg algebra, in~which classical dynamics are regarded as the  
$\hbar\rightarrow 0$ limit of quantum theory, 
\begin{eqnarray} \label{xpPB1-A}
\lim_{\hbar\rightarrow 0}\frac{1}{i\hbar} \left[\hat{x}^i , \hat{p}_j\right]&:=& \left\{x^i , p_j \right\}=\delta^{i}{}_{j}, 
\end{eqnarray}
\begin{eqnarray} \label{xpPB1-B}
\lim_{\hbar\rightarrow 0}\frac{1}{i\hbar} \left[\hat{x}^i , \hat{x}^j\right]&:=& \left\{x^i , x^j \right\}=0, 
\end{eqnarray}
\begin{eqnarray} \label{xpPB1-C}
\lim_{\hbar\rightarrow 0}\frac{1}{i\hbar} \left[\hat{p}_i , \hat{p}_j\right]&:=& \left\{p_i , p_j \right\}=0.
\end{eqnarray}
Note that here the~$x^i$ denote global Cartesians, not generalised coordinates, and~the $p_i$ denote the corresponding linear momenta~\cite{Takhtajan1}.
For a classical observable $O({\bf x},{\bf p})$, the~corresponding quantum operator is then defined as $\hat{O}(\hat{{\bf x}},\hat{{\bf p}})$, up~to ordering ambiguities caused by the noncommutativity of the position and momentum~operators.

In general, any pair of operators $\hat{f},\hat{g}$ preserves the canonical structure
\begin{equation}\label{fgCM}
\lim_{\hbar\rightarrow 0}\left[\hat{f},\hat{g}\right]=\left\{f,g\right\}.
\end{equation}
The position--momentum commutator defined by canonical quantization therefore inherits all the familiar properties of the Poisson bracket, namely, skew-symmetry, bilinearity, and~compatibility with the Leibniz rule and the Jacobi identities (\ref{PSBK2-A})-(\ref{PSBK2-E}).

\subsection{From Heisenberg to the Lie, Clifford, and~Dirac~Algebras} \label{Sec.3.3}

Based on the canonical quantization procedure, the~angular momentum operator is defined as $\hat{L}_i=i\epsilon_{ij}{}^{k}\hat{x}^j\hat{p}_k$.
The components of the angular momentum obey the Lie algebra~\cite{Takhtajan1}
\begin{equation}\label{XY2}
[\hat{L}_i,\hat{L}_j] = i\hbar\epsilon_{ij}{}^{k}\hat{L}_{k},
\end{equation}
where $\epsilon_{ij}{}^{k}$ is the three-dimensional Levi-Civita symbol. 
Similarly, the Pauli matrices
\begin{equation}\label{sima1}
\sigma_1=\left(
\begin{array}{cc}
0 & 1 \\
1 & 0
\end{array}\right), \quad
\sigma_2=\left(
\begin{array}{cc}
0 & -i \\
i & 0
\end{array}\right), \quad
\sigma_2=\left(
\begin{array}{cc}
1 & 0 \\
0 & -1
\end{array}\right),
\end{equation}
which are used to define the spin-1/2 operators $\hat{s}_i = (\hbar/2)\sigma_i$, also obey the Lie algebra
\begin{equation}\label{SS1}
\left[\sigma_i,\sigma_j\right] = 2i\epsilon_{ij}{}^{k}\sigma_{k},
\end{equation}
as well as the Clifford algebra
\begin{equation}\label{SS1}
\left\{\sigma_i,\sigma_j\right\} = 2i\delta_{ij}\mathbb{I},
\end{equation}
where $\left\{\sigma_i,\sigma_j\right\} :=\sigma_i\sigma_j+\sigma_j\sigma_i$ again denotes the anticommutator~\cite{Armin}.

To describe both particles and antiparticles, Dirac introduced a set of operators known as the gamma matrices, which are often denoted by using the slightly inconsistent notation
$(\gamma^{\mu},\gamma^5) = (\gamma^0,\gamma^i,\gamma^5)$, where $\mu \in \left\{0,1,2,3\right\}$ and $i \in \left\{1,2,3\right\}$.
Explicitly, these may be written as
\begin{equation}\label{gama1}
\gamma^0=\left(
\begin{array}{cc}
I & 0 \\
0 & -I
\end{array}\right), \quad
\gamma^i=\left(
\begin{array}{cc}
0 & \sigma_i \\
-\sigma_i & 0
\end{array}\right),
\quad
\gamma^5=\left(
\begin{array}{cc}
0 & I \\
I & 0
\end{array}\right),
\end{equation}
and it is straightforward to demonstrate that $\gamma^5:=i\gamma^0\gamma^1\gamma^2\gamma^3$.
The Dirac operators obey the following relations~\cite{Armin} 
\begin{eqnarray} \label{DA1-A}
\left\{\gamma^\mu,\gamma^\nu\right\} &=& 2\eta^{\mu\nu} \mathbb{I} 
 \quad \textrm{(Clifford algebra)}, 
 \end{eqnarray}
\begin{eqnarray} \label{DA1-B}
\left\{\gamma^5,\gamma^\nu\right\} &=& 0, 
\end{eqnarray}
\begin{eqnarray} \label{DA1-C}
\left\{\gamma^5,\sigma^{\mu\nu}\right\} &=& 0,
\end{eqnarray}
\begin{eqnarray} \label{DA1-D}
\gamma^\mu\gamma^\nu &=& (\eta^{\mu\nu}-i\sigma^{\mu\nu})\mathbb{I}, 
\end{eqnarray}
\begin{eqnarray} \label{DA1-E}
\left(\gamma^\mu\right)^2 &=& \eta^{\mu\mu}\mathbb{I}, 
\end{eqnarray}
\begin{eqnarray} \label{DA1-F}
\gamma_\mu\gamma^\mu &=& 4,
\end{eqnarray}
where $\eta^{\mu\nu}=(1,-1,-1,-1)$ is the Minkowki metric and $\sigma^{\mu\nu}:=\frac{i}{2}[\gamma^\mu,\gamma^\nu]$.
These \linebreak{(anti-)commutation} relations are called the Dirac~algebra.

\subsection{The Snyder~Algebra} \label{Sec.3.4}

To regularise the emergence of singularities in particle physics, while preserving Lorentz invariance, Snyder introduced a noncommutative 
algebra formulated in five-dimensional spacetime with a constraint, namely 
\begin{eqnarray} \label{SYder1-A}
x &=& ia \left(\eta^4\frac{\partial}{\partial\eta^1}-\eta^1\frac{\partial}{\partial\eta^4}\right), 
\end{eqnarray}
\begin{eqnarray} \label{SYder1-B}
y &=& ia \left(\eta^4\frac{\partial}{\partial\eta^2}-\eta^2\frac{\partial}{\partial\eta^4}\right), 
\end{eqnarray}
\begin{eqnarray} \label{SYder1-C}
z &=& ia \left(\eta^4\frac{\partial}{\partial\eta^3}-\eta^3\frac{\partial}{\partial\eta^4}\right), 
\end{eqnarray}
\begin{eqnarray} \label{SYder1-D}
t &=& \frac{ia}{c}\left(\eta^4\frac{\partial}{\partial\eta^0}+\eta^0\frac{\partial}{\partial\eta^4}\right),
\end{eqnarray}
where $a$ is the noncommutative parameter, the~$\eta_i$ are real variables, and~the constraint
\begin{equation}\label{DSTer}
-\eta^2=\eta_0^2-\eta_1^2-\eta_2^2-\eta_3^2-\eta_4^2
\end{equation}
defines the embedding of a four-dimensional de Sitter geometry~\cite{Snyder}.
Consequently, the~spacetime variables obey the following commutation relations,
\begin{eqnarray} \label{SYder2-A}
[x^i,x^j] &=& \frac{ia^2}{\hbar}\epsilon^{ijk}L_k, 
\end{eqnarray}
\begin{eqnarray} \label{SYder2-B}
[t,x^i] &=& \frac{ia^2}{\hbar c}M^i,
\end{eqnarray}
where 
\begin{eqnarray} \label{SYder3-A}
L_i &=& i\hbar \epsilon_{i}{}^{jk}\eta_j\frac{\partial}{\partial\eta^k}, 
\end{eqnarray}
\begin{eqnarray} \label{SYder3-B}
M^i &=& i\hbar \left(\eta^0\frac{\partial}{\partial\eta_i}+\eta^i\frac{\partial}{\partial\eta_0}\right).
\end{eqnarray}
When $a\rightarrow 0$ the Snyder algebra (\ref{SYder2-A})-(\ref{SYder2-B}) reduces to the standard commutative relations of ordinary Minkowski space.
This noncommutative algebra preserves Lorentz invariance and can help to alleviate or avoid singularities in quantum field theories, defined on the associated background geometry, without~the use of renormalization techniques~\cite{Snyder}.

\subsection{Beyond Poisson: The Nambu~Algebra} \label{Sec.3.5}

Nambu's original idea was to look for an alternative formalism for Hamiltonian mechanics, which preserves the volume of the phase space (the Liouville theorem), and which, therefore, can be applied to statistical ensembles~\cite{Nambu}.
To achieve this, he introduced the so-called $N$-triplet canonical variables, and~an ``extended'' Poisson bracket.
The Nambu formalism of extended Hamiltonian mechanics was later quantized, giving rise to the so-called ``Nambu quantum mechanics'' \cite{Curtright,Sahoo}, which can be used, among~other applications, to~describe quark triplets~\cite{Sucre}, magnetic monopoles~\cite{Hirayama1}, and~superconductivity~\cite{Angulo}.
The Nambu dynamics also inspired various mathematicians, who later reformulated the theory in geometric terms~\cite{Fecko}.
Here, we introduce the basic formulation of Nambu dynamics and briefly describe the procedure for quantizing classical Nambu systems.

In Nambu mechanics, the~Poisson pair of canonical variables in two-dimensional phase space, $(x,p)$, is extended to a triplet of dynamical variables in a three-dimensional phase space, $\mathbf{x}=(x^1,x^2,x^3)$.
Two Hamiltonian functions are introduced on this space, denoted as $H_1$ and $H_2$, yielding a generalized Hamilton equation~\cite{Nambu,Bayen,Estabrook},
\begin{equation}\label{GHE1}
\frac{d\mathbf{x}}{dt}=\nabla H_1\times\nabla H_2.
\end{equation}

The divergence of the velocity field, ${\bf v} = d{\bf x}/dt$, is given by $\nabla\cdot \mathbf{v}=\nabla\cdot (\nabla H_1\times\nabla H_2)$.
The invariance of the phase space volume requires $\nabla\cdot (\nabla H_1\times\nabla H_2)=0$, which corresponds to a generalized Liouville theorem.
Similarly, for~any function $F(x^1,x^2,x^3)$, the~equation of motion is given by
\begin{equation}\label{GHE2}
\frac{dF}{dt}=\nabla F\cdot (\nabla H_1\times\nabla H_2),
\end{equation}
where
\begin{equation}\label{GHE3}
\nabla F\cdot (\nabla H_1\times\nabla H_2)=\epsilon^{ijk}\partial_iF\partial_jH_1\partial_kH_2.
\end{equation}

The Nambu bracket, which is a generalization of the canonical Poisson bracket, is defined as
\begin{equation}\label{NB1}
\left\{A_1,A_2,A_3\right\}:=\epsilon^{ijk}\partial_iA_1\partial_jA_2\partial_kA_3.
\end{equation}

It possesses the following properties and obeys the following relations.

\begin{enumerate}

  \item Skew-symmetry:  \begin{equation}\label{SS1}
  \left\{A_1,A_2,A_3\right\}=(-1)^{\epsilon(p)}\left\{A_{p(1)},A_{p(2)},A_{p(3)}\right\},
  \end{equation}
   where $p(i)$ is the permutation of indices and $\epsilon(p)$ is the parity of the~permutation.
   
  \item Leibniz rule: \begin{equation}\label{DS1}
  \left\{A_1A_2,A_3,A_4\right\}=A_1\left\{A_2,A_3,A_4\right\}+\left\{A_1,A_3,A_4\right\}A_2.
  \end{equation}
  
  \item Fundamental identity: 

  \begin{equation}\label{FI1}
  \left\{A_1,A_2,\left\{A_3,A_4,A_5\right\}\right\} = \nonumber
  \end{equation}
  \begin{equation}
  \left\{\left\{A_1,A_2,A_3\right\},A_4,A_5\right\}+\left\{A_3,\left\{A_1,A_2,A_4\right\},A_5\right\}+\left\{A_3,A_4,\left\{A_1,A_2,A_5\right\}\right\}.
  \end{equation}
 
\end{enumerate}

In particular, one has that
\begin{equation}\label{NB2}
\left\{x^i,x^j,x^k\right\}=\epsilon^{ijk}
\end{equation}
and
\begin{equation}\label{NB3}
\left\{A_j,A_j,A_i\right\}=\left\{A_j,A_i,A_j\right\}=\left\{A_i,A_j,A_j\right\}=0,
\end{equation}
for any $i,j \in \left\{1,2\right\}$.
The corresponding generalisation of Hamiltonian mechanics on a $2d$-dimensional symplectic manifold, with~canonical variables $(x^i,p_j)$, $i,j \in \left\{1,2, \dots, d\right\}$ is straightforward, but~the relations are cumbersome to write, and we neglect them here for the sake of pedagogical~clarity.

The Nambu bracket structure is preserved under differential maps $x^i\rightarrow y^i(x)$, which preserve the volume of the phase space, such that $\left\{y^1,y^2,y^3\right\}=1$, i.e.,~$\left\{y^1,y^2,y^3\right\}:=\epsilon^{ijk}\partial_iy^1\partial_jy^2\partial_ky^3$.
These are called the volume-preserving diffeomorphisms (VPD).
{In general, these involve two independent functions, $f$ and $g$, and~the infinitesimal three-dimensional generator of VPDs is defined as~\cite{Awata}}
\begin{equation}\label{VPDG1}
D(f,g):=\epsilon^{ijk}\partial_if\partial_jg\partial_k\equiv D^k(f,g)\partial_k.
\end{equation}
The volume-preserving property implies $\partial_i D^i(f,g)=\partial_k\left(\epsilon^{ijk}\partial_i f\partial_j g\right)=0$, which is equivalent to the divergencelessness of the velocity $\nabla\cdot \mathbf{v}=0$.

We may also consider parametrizations of the triple phase space, of~the form $\left\{x^\alpha\right\}\rightarrow \left\{X^\alpha\right\}$, where $\alpha=0,1,\cdots,d$. The~induced infinitesimal volume element is given by~\cite{Awata}
\begin{equation}\label{VE1}
d\sigma=\sqrt{\left\{X^\alpha,X^\beta,X^\gamma\right\}}dx^1dx^2dx^3,
\end{equation}
where the volume element is invariant under the transform $Y=X+\epsilon D(f,g)X$.
The Nambu bracket is also invariant under this transformation,
\begin{eqnarray}\label{VE2}
\left\{Y^\alpha,Y^\beta,Y^\gamma\right\}-\left\{X^\alpha,X^\beta,X^\gamma\right\}&=&\epsilon D(f,g)\left\{X^\alpha,X^\beta,X^\gamma\right\} 
+\mathcal{O}(\epsilon^2).
\end{eqnarray}

The quantization of the classical Nambu brackets in Equation (\ref{NB1}) should preserve the properties in Equations (\ref{SS1})--(\ref{FI1}).
Following the canonical quantization procedure, the~quantum Nambu bracket is defined as
\begin{equation}\label{QNB1}
\lim_{\hbar\rightarrow 0}\frac{1}{i\hbar}\left[A_1,A_2,A_3\right]=\left\{A_1,A_2,A_3\right\},
\end{equation}
such that it obeys the the following properties~\cite{Curtright,Awata}.
\begin{enumerate}
  \item Skew-symmetry:
\begin{equation}\label{QSS1}
  \left[A_1,A_2,A_3\right]=(-1)^{\epsilon(p)}\left[A_{p(1)},A_{p(2)},A_{p(3)}\right],
  \end{equation}
   where $p(i)$ is the permutation of indices and $\epsilon(p)$ is the parity of the permutation.
  \item Leibniz rule:
\begin{equation}\label{QDS1}
  \left[A_1A_2,A_3,A_4\right]=A_1\left[A_2,A_3,A_4\right]+\left[A_1,A_3,A_4\right]A_2.
  \end{equation}
  \item Fundamental identity: 
\begin{equation}\label{QFI1}
  \left[A_1,A_2,\left[A_3,A_4,A_5\right]\right]=
  \left[\left[A_1,A_2,A_3\right],A_4,A_5\right]+\left[A_3,\left[A_1,A_2,A_4\right],A_5\right]+\left[A_3,A_4,\left[A_1,A_2,A_5\right]\right].
  \end{equation}
\end{enumerate}

In particular,
\begin{equation}\label{QNB2}
\left[x^\alpha,x^\beta,x^\gamma\right]=i\hbar\epsilon^{\alpha\beta\gamma},
\end{equation}
and
\begin{equation}\label{NB3}
\left[A_{\beta},A_{\beta},A_\alpha\right]=\left[A_{\beta},A_\beta,A_{\beta}\right]=\left[A_\alpha,A_{\beta},A_{\beta}\right]=0,
\end{equation}
for all $\alpha,\beta$.
Written explicitly, the~quantum Nambu bracket can be expressed in terms of the Heisenberg canonical bracket as~\cite{Awata,Sahoo,Zachos}
\begin{equation}\label{NBPB1}
\left[A,B,C\right]=A\left[B,C\right]+B\left[C,A\right]+C\left[A,B\right].
\end{equation}

The quantum Nambu equation is given by
\begin{equation}\label{QHBE1}
i\hbar\frac{dF}{dt} =\left[F,H_1,H_2\right]
\end{equation}
for an arbitrary function $F$.

When the triple variables are generalized to $n$-component vectors $x^i$ and $n-1$ Hamiltonian functions, the~$n$-component Nambu bracket also keeps the above properties.
However, it is still unclear how to link Nambu mechanics to Hamiltonian and Lagrangian mechanics, directly, and what relationship should exist between the extended and canonical variables.
Most importantly, it remains unclear how to construct multiple Hamiltonian functions on a phase space with an odd number of dimensions~\cite{Bayen,Mukunda}.

\subsection{The Deformed Heisenberg Algebra: Noncommutative Phase~Space} \label{Sec.3.6}

Various generalizations of the Heisenberg commutation relations have been considered in the existing literature on noncommutative geometry, and~these have been based on various physical arguments.
In this Section, we review some of the best explored and motivated proposals; for more details, see \cite{Douglas,Szabo,Konechny,Rosenbaum,Gouba,Francois,Kovacik} and references therein.

Model I: Gedanken experiments in phenomenological quantum gravity, as~well as several specific approaches to this problem, including string theory, loop quantum gravity, and~others, suggest the existence of a minimum resolvable length scale in nature, of~the order of the Planck length~\cite{Szabo,Carlo}.
The minimal implementation of this idea therefore suggests that the spatial coordinates may be noncommutative but~that the canonical 
Heisenberg commutation relations between the positions and momenta are preserved, giving~\cite{Masud,Basu}
\begin{eqnarray}
[\hat{X}^i,\hat{P}_j] &=& i\hbar\delta^{i}{}_{j}\hat{\mathbb{I}}, \label{XYPP1a}
\end{eqnarray}
\begin{eqnarray}
[\hat{X}^i,\hat{Y}^j] &=& i\theta^{ij}\hat{\mathbb{I}}, \label{XYPP1b}
\end{eqnarray}
\begin{eqnarray}
[\hat{P}_i,\hat{P}_j] &=& 0. \label{XYPP1c} 
\end{eqnarray}

The noncommutative relations (\ref{XYPP1b}) yield additional quantum fluctuations between different spatial components, giving rise to the 
generalized uncertainty relations $\Delta X^i\Delta Y^j\geq \theta^{ij}/2$. 

Model II: In some models, the~existence of dark energy, which drives the present-day accelerated expansion of the universe, is associated with the existence of a minimum resolvable momentum, and~hence energy~\cite{Sabino,Peebles}.
The physical scenario considered here is one in which there exists a minimum energy density, and~hence a minimum curvature of spacetime, which arises directly from quantum fluctuations of the canonical momentum components, 
yielding~\cite{Fredenhagen}
\begin{eqnarray}
[\hat{X}^i,\hat{P}_j] &=& i\hbar\delta^{i}{}_{j}\hat{\mathbb{I}}, \label{XYPP2a}
\end{eqnarray}
\begin{eqnarray}
[\hat{X}_i,\hat{Y}_j] &=& 0, \label{XYPP2b}
\end{eqnarray}
\begin{eqnarray}
[\hat{P}_i,\hat{P}_j] &=& i\eta_{ij}\hat{\mathbb{I}}. \label{XYPP2c} 
\end{eqnarray}

The noncommutative relations (\ref{XYPP2c}) yield additional quantum fluctuations between different momenta, such 
that $\Delta P_i\Delta P_j\geq \eta_{ij}/2$. 

Model III: Combining both arguments above motivates us to include both spatial and momentum noncommutative relations, while still leaving the 
canonical position--momentum commutator of the Heisenberg algebra intact, giving~\cite{Liang,Liang1,Harko}
\begin{eqnarray}
[\hat{X}^i,\hat{P}_j] &=& i\hbar\delta^{i}{}_{j}\hat{\mathbb{I}}, \label{XYPP31}
\end{eqnarray}
\begin{eqnarray}
[\hat{X}^i,\hat{Y}^j] &=& i\theta^{ij}\hat{\mathbb{I}}, \label{XYPP3b}
\end{eqnarray}
\begin{eqnarray}
[\hat{P}_i,\hat{P}_j] &=& i\eta_{ij}\hat{\mathbb{I}}. \label{XYPP3c} 
\end{eqnarray}

In this case, we obtain additional quantum fluctuations arising from the noncommutativity of both the spatial and momentum~components.

Model IV: The physical scenario considered in Model III can be generalized even further, to~modify the canonical Heisenberg 
relation between position and momentum, such that~\cite{Liang,Liang1,Harko}
\begin{eqnarray}
[\hat{X}^i,\hat{P}_j] &=& i\kappa^{i}{}_{j}\hat{\mathbb{I}}, \label{XYPP4a} 
\end{eqnarray}
\begin{eqnarray}
[\hat{X}^i,\hat{Y}^j] &=& i\theta^{ij}\hat{\mathbb{I}}, \label{XYPP4b}
\end{eqnarray}
\begin{eqnarray}
[\hat{P}_i,\hat{P}_j] &=& i\eta_{ij}\hat{\mathbb{I}}, \label{XYPP4c}
\end{eqnarray}
where $\kappa^{i}{}_{j} \neq \hbar\delta^{i}{}_{j}$. 
We discuss this generalization in detail in Section \ref{Sec.4} below. 

However, before concluding this Section, let us note that each of the models above can also be motivated in a number of different ways.
For example, models, in which $[\hat{P}_i,\hat{P}_j] \neq 0$ can be derived from the theory of relative locality~\cite{Giovanni*}, where the noncommutativity of the momentum components is generated via the curvature of momentum space. 
Noncommutative generalizations of the canonical Heisenberg algebra can also be derived by considering deformations of the canonical spacetime symmetries, as~in the $\kappa-$Poincare approach to noncommutative geometry, which has been extensively applied in the construction of noncommutative field theories~\cite{Giovanni}.
These results suggest deep geometric connections between the deformed phase space, noncommutative algebras, and quantum~geometry.}

\section{Noncommutative Quantum~Mechanics} \label{Sec.4}

In this Section, we give a pedagogical introduction to one of the most studied methods, used to generalize the canonical Heisenberg algebra to noncommutative phase space, namely, the SW map~\cite{Gouba,Sivasubramanian,Kokado,Vilela,Harko}.

\subsection{Position and Momentum in Noncommutative Phase~Space} \label{Sec.4.1}

In general, the~canonical Heisenberg phase space can be generalized to a noncommutative phase space by imposing the algebra (\ref{XYPP4a})-(\ref{XYPP4c}).
For simplicity, one may assume that all nonzero components $\theta^{ij}$, $\eta_{ij}$ are equal in magnitude, i.e.,~$\theta^{ii} = 0$, $\eta_{ii} = 0$ and $\theta^{ij} = -\theta^{ji} = \theta$, $\eta_{ij} = -\eta_{ji} = \eta$, for~$ i\neq j$.
Similarly, we assume that the diagonal components of the matrix $\kappa^{i}{}_{j}$ are equal, $\kappa^{i}{}_{i} = \kappa^{\alpha}$, and~that the 
off-diagonal components differ by at most a change of sign, $\kappa^{i}{}_{j} = - \kappa^{j}{}_{i} = \kappa^{\beta}$, as~required by the symmetries of the noncommutative \mbox{relations (\ref{XYPP4a})-(\ref{XYPP4c})}.
For later convenience, we write down the noncommutative matrices, explicitly, as

\begin{equation}  \label{ABC}
[\theta^{ij}]= \left(
\begin{array}{ccc}
0 & \theta & \theta \\
-\theta & 0 & \theta \\
-\theta & -\theta & 0
\end{array}
\right), \quad
[\eta_{ij}] =\left(
\begin{array}{ccc}
0 & \eta & \eta \\
-\eta & 0 & \eta \\
-\eta & -\eta & 0
\end{array}
\right) , \quad
[\kappa^{i}{}_{j}] =\left(
\begin{array}{ccc}
\kappa^\alpha & \kappa^\beta & -\kappa^\beta \\
\kappa^\beta & \kappa^\alpha & \kappa^\beta \\
-\kappa^\beta & \kappa^\beta & \kappa^\alpha
\end{array}
\right).
\end{equation}

Note that here we choose to parameterize the antisymmetric matrices $\theta^{ij}$ and $\eta_{ij}$ and~the symmetric matrix $\kappa^{i}{}_{j}$ with~the minimum possible number of independent parameters, i.e.,~$\theta$, $\eta$, $\kappa^{\alpha}$ and $\kappa^{\beta}$.
However, this choice is made for the sake of simplicity and pedagogical clarity, and~more complicated models, with~more parameters, may be more suitable for the description of various physical scenarios.
In the Section \ref{Sec.4.8} below, one can see how this ``minimal'' set of  noncommutative parameters can be naturally associated with universal physical constants, such as the Planck length and cosmological constant, which may provide a physical scenario for dark energy and quantum gravity, in~terms of noncommutative geometry.
Moreover, one can see that the $\kappa^{i}{}_{j}$ matrix parameters, $\kappa^\alpha$ and $\kappa^\beta$, can be naturally expressed in terms of $\theta$ and $\eta$, based on the 
SW map with the Bopp shift~\cite{Bertolami1}.

The corresponding generalised uncertainty relations are 
\begin{eqnarray}
\Delta X^i\Delta P_j\geq \frac{\kappa^{i}{}_{j}}{2}, \label{GUR-1a}
\end{eqnarray}
\begin{eqnarray}
\Delta X^i\Delta Y^j\geq \frac{\theta^{ij}}{2}, \label{GUR-1b}
\end{eqnarray}
\begin{eqnarray}
\Delta P_i\Delta P_j\geq \frac{\eta_{ij}}{2}. \label{GUR-1c}
\end{eqnarray}

\subsection{Angular Momenta in Noncommutative Phase~Space} \label{Sec.4.2}

In the canonical quantum mechanics, the~angular momentum operators are the generators of the rotation group $SO(3)$.
A vector rotation is expressed as
\begin{equation}\label{UU1}
\hat{\bf{x}}'= \hat{\bf{x}}+\delta \boldsymbol{\phi}\times \hat{\bf{x}}, 
\end{equation}
where $\delta\boldsymbol{\phi}=(\delta \phi_x, \delta\phi_y,\delta\phi_z)$ are the infinitesimal Euler anglers. 
The wave function transforms under rotations according to
\begin{equation}\label{RWF}
\psi'(\bf{\bf{x}},t)= \hat{U}_R(\delta\boldsymbol{\phi})\psi(\hat{\bf{x}},t), 
\end{equation}
where
\begin{equation}\label{UU2}
\hat{U}_R(\delta\boldsymbol{\phi})=I-\frac{i}{\hbar}\delta\boldsymbol{\phi} \cdot \hat{\bf{\bf{L}}}, 
\end{equation}
and $\hat{\bf{L}}=\hat{\bf{x}}\times \hat{\bf{p}}$ is the operator counterpart of the orbital angular momentum pseudovector.
The components of the angular momentum thus satisfy the closed relations $[\hat{L}_\alpha,\hat{L}_\beta]=i\hbar \varepsilon_{\alpha\beta}{}^{\gamma}\hat{L}_\gamma$, which are equivalent to the $\mathfrak{su(2)}$ Lie algebra.
In isotropic space, $[\hat{U}_R(\boldsymbol{\phi}),\hat{H}]=0$ and $[\hat{L},\hat{H}]=0$, which implies that the angular momentum is conserved.

In the noncommutative phase space, we assume the rotation transformations (\ref{UU1})--(\ref{UU2}) still hold, but~$\hat{\bf{x}}\rightarrow \hat{\bf{X}}$, $\hat{\bf{p}}\rightarrow \hat{\bf{P}}$ and 
$\hat{\bf{L}}\rightarrow \hat{\bf{L}}=\hat{\bf{X}}\times \hat{\bf{P}}$. 
In~other words, the angular momentum operator in the noncommutative phase space is given explicitly as
\begin{equation}\label{AM1}
\hat{L}_i=\epsilon_{ij}{}_{k}\hat{X}^{j}\hat{P}_{k}.
\end{equation}
By using the basic noncommutative relations in (\ref{XYPP4a})-(\ref{XYPP4c}), the~commutation relations for the components of angular momenta can be obtained as
\begin{equation}\label{AMNCR1}
[\hat{L}_\alpha,\hat{L}_\beta]=
i\epsilon_{\alpha j}{}^{k}\epsilon_{\beta b}{}^{c}[(\kappa^{j}{}_{c}\hat{X}^{b}\hat{P}_{k}-\kappa^{b}{}_{k}\hat{X}^{j}\hat{P}_{c})
+(\eta_{kc}\hat{X}^{j}\hat{X}^{b}+\theta^{jb}\hat{P}_{c}\hat{P}_{k})].
\end{equation}
It can be seen that the angular momentum does not obey the $SO(3)$ Lie algebra.
It can be verified that $[\hat{L},\hat{H}] \neq 0$,
which implies that rotational symmetry is broken in the noncommutative phase~space.

However, it should be noted that in~some approaches to noncommutative geometry, such as the $\kappa-$Minkowski model of noncommutative spacetime, generalized operators are taken, by~definition, as~the generators of deformed-symmetry transformations~\cite{Alessandra,Michele}.
In this approach, the~Lie algebras of canonical quantum mechanics are replaced by Hopf algebras, and~the dynamics of the canonical theory are generalized to include additional couplings between matter and gravity, as~required by the structure of the $\kappa-$Poincare Hopf algebra~\cite{Alessandra,Michele}.

\subsection{The Heisenberg Equation and Conservation~Laws} \label{Sec.4.3}

As in canonical quantum mechanics, it is assumed that the equation of motion for a time-dependent 
operator, $\hat{Q}$, in~the noncommutative phase space, is the Heisenberg~equation
\begin{equation}\label{EqM}
\frac{d}{dt} \hat{Q}=\frac{1}{i\hbar}\left[\hat{Q},\hat{H}\right],
\end{equation}
where $[\hat{Q},\hat{H}]$ is the appropriate (perhaps noncanonical) commutator.

According to Noether's theorem, a~physical variable $\hat{Q}$ is conserved under a unitary transformation, $\hat{U}=e^{-i\alpha\widehat{Q}}$, where $\alpha\in \mathbb{R} $ is a continuous parameter, if~$[\hat{Q},\hat{H}]=0$.
In canonical quantum mechanics, the~spatial translation operator, which acts according to $\hat{T}_\mathbf{X}:\hat{\mathbf{X}}\rightarrow \hat{\mathbf{X}}+\mathbf{a}$, $\psi\left(\hat{\mathbf{X}},t\right)\rightarrow \hat{T}_\mathbf{X}\psi\left(\hat{\mathbf{X}},t\right)=\psi\left(\hat{\mathbf{X}}+\mathbf{a},t\right)$, is given by $\hat{T}_\mathbf{X}=e^{\frac{i}{\hbar}\mathbf{a}\cdot \hat{\mathbf{P}}}$.
Hence, the~momentum operator can be regarded as the generator of the translation group. 
However, in~the noncommutative phase space, we have $[\hat{T}_\mathbf{X},\hat{H}]\neq 0$ and $[\hat{\mathbf{P}},\hat{H}]\neq 0$, even for the free particle with 
Hamiltonian, $\hat{H}=\hat{\mathbf{P}}^{2}/(2m)$.
As with the generalized angular momentum operators, discussed in Section~\ref{Sec.4.3}, one can just define generalised ``translations'' as a group of transformations in the noncommutative phase space, then identify their generators with the (noncommutative) components of the generalized momenta (c.f.~\cite{Alessandra,Michele}).

In addition, assuming that the vector momentum operator is given in the usual way, $\hat{\mathbf{P}} = \hat{P}_X{\bf i} + \hat{P}_Y{\bf j} + \hat{P}_Y{\bf k}$, and~using the basic noncommutative relations in (\ref{XYPP4a})-(\ref{XYPP4c}), we obtain
\begin{equation}\label{PHCR}
[\hat{\mathbf{P}},\hat{H}]=\frac{\eta}{m}\hat{\mathbf{K}}_p,
\end{equation}
where
\begin{equation}\label{2Pr1}
\hat{\mathbf{K}}_p=\left(\hat{P}_y+\hat{P}_z\right)\mathbf{i}-\left(\hat{P}_x-\hat{P}_z\right)\mathbf{j}.
-\left(\hat{P}_x+\hat{P}_y\right)\mathbf{k}.
\end{equation}

The noncommutative relation in (\ref{PHCR}), between~the momentum and Hamiltonian operators, implies that $\frac{d}{dt} \left\langle\hat{\mathbf{P}}\right\rangle = \frac{\eta}{m}\left\langle\hat{\mathbf{K}}_p\right\rangle\neq 0$.
Namely, that momentum is not conserved in the noncommutative phase space, even for a free particle.
This is because spatial translational symmetry is broken.
However, interestingly, it can be verified that $[\hat{\mathbf{P}}^2,\widehat{H}]=0$, which implies that the amplitude of the momentum is still conserved.
In other words, the~direction of the momentum shows an intrinsic stochastic behavior, in~the noncommutative phase space, due to fluctuations of the background geometry, but~this does not alter the total energy of the~system.

The commutation relation between the angular momentum operator and the Hamiltonian of the free particle is obtained as
\begin{equation}\label{LHCR}
\left[\hat{\mathbf{L}},\hat{H}\right]
=\frac{i\hbar}{m}\left(\hat{\Omega}_X\mathbf{i}+\hat{\Omega}_Y\mathbf{j}+\hat{\Omega}_Z\mathbf{k}\right)\equiv\frac{i\hbar}{m}\hat{{\Omega}},
\end{equation}
where 
\begin{eqnarray}\label{LHCR2}
\widehat{\Omega}_X &=& \frac{\eta}{\hbar}\left(-\left\{\hat{Y},\hat{P}_{Y}\right\}-\left\{\hat{Z},\hat{P}_Z\right\}
-\left\{\hat{Y},\hat{P}_X\right\}+\left\{\hat{Z},\hat{P}_X\right\}\right)
\nonumber\\
&+&\frac{\kappa^\beta}{\hbar}\left(\left\{\hat{P}_X,\hat{P}_Y\right\}+\left\{\hat{P}_X,\hat{P}_Z\right\}+2 \hat{P}_{Y}^{2}+2\hat{P}_{Z}^{2}\right), 
\end{eqnarray}
\begin{eqnarray}
\widehat{\Omega}_Y &=&
\frac{\eta}{\hbar}\left(\left\{\hat{X},\hat{P}_{X}\right\}+\left\{\hat{X},\hat{P}_Y\right\}+\left\{\hat{Z},\hat{P}_Z\right\}+\left\{\hat{Z},\hat{P}_Y\right\}\right)
\nonumber\\
&+&\frac{\kappa^\beta}{\hbar} \left(\left\{\hat{P}_X,\hat{P}_Y\right\}-\left\{\hat{P}_Y,\hat{P}_Z\right\}- 2\hat{P}_{X}^{2}+2\hat{P}_{Z}^{2}\right), 
\end{eqnarray}
\begin{eqnarray}
\widehat{\Omega}_Z &=&
\frac{\eta}{\hbar}\left(-\left\{\hat{X},\hat{P}_{X}\right\}-\left\{\hat{Y},\hat{P}_Y\right\}+\left\{\hat{X},\hat{P}_Z\right\}-\left\{\widehat{Y},\widehat{P}_z\right\}\right)
\nonumber\\
&+&\frac{\kappa^\beta}{\hbar} \left(-\left\{\hat{P}_X,\hat{P}_Z\right\}-\left\{\hat{P}_Y,\hat{P}_Z\right\}- 2\hat{P}_{X}^{2}+2\hat{P}_{Y}^{2}\right). 
\end{eqnarray}
In general, for~$\eta \neq 0$ and $\kappa^\beta \neq 0$, we have $\hat{{\Omega}}\neq 0$.
This implies that $\frac{d}{dt} \left\langle\hat{\mathbf{L}}\right\rangle = \frac{i\hbar}{m}\left\langle\hat{{\Omega}}\right\rangle\neq 0$, i.e.,~that the angular momentum is not  conserved in noncommutative phase space, even for free particles.
However, whether there exist specific states, and/or specific nonzero values of $\eta$ and $\kappa^{\alpha}$, such that $\langle\hat{{\Omega}}\rangle =0$, is an interesting question, although~we 
do not attempt to answer it here. 
In addition, the commutation relation between the square of the angular momentum and the Hamiltonian can be 
expressed as
\begin{equation}\label{LHCR2}
[\hat{\mathbf{L}}^2,\hat{H}]=\frac{i\hbar}{m}\left(\hat{\mathbf{L}}\cdot\hat{{\Omega}}+\hat{{\Omega}}\cdot\hat{\mathbf{L}}\right),
\end{equation}
which implies that the amplitude of the angular momentum is also not conserved, in~contrast to the energy.
In general, both spatial translation and rotation symmetries are broken in the noncommutative phase~space.

By contrast, time translations are defined implicitly through the Heisenberg equation, as~the transformations generated by the Hamiltonian, $\hat{H}$.
Hence, the~preservation of time-translation symmetry is a direct consequence of the Heisenberg equation, which is assumed to also hold in~the noncommutative phase space model.
More specifically, the~time-translation operator, which acts according to $\hat{T}_t:t\rightarrow t+\tau$ and $\psi(\hat{\mathbf{X}},t)\rightarrow \hat{T}_t\psi(\hat{\mathbf{X}},t)=\psi(\hat{\mathbf{X}},t+\tau)$, is given by $\hat{T}_t=e^{\tau\frac{\partial}{\partial t}}=e^{-\frac{i}{\hbar}\tau \hat{H}}$, where $\hat{H}=i\hbar \frac{\partial}{\partial t}$.
We then have $\left[\hat{T}_t,\hat{H}\right]=0$ and $[\hat{H},\hat{H}]=0$, i.e.,~time-translation symmetry still holds in the noncommutative phase space, which corresponds to the conservation of energy.

However, since spatial translation symmetry is broken in the noncommutative phase space, let us investigate the velocity and acceleration of a free particle.
The velocity of the particle is defined as $\hat{\mathbf{v}}:=\frac{d}{dt} \hat{\mathbf{X}}$, so that by using the Heisenberg 
equation~(\ref{EqM}) one has: 
\begin{equation}\label{Vp}
\hat{\mathbf{v}}=\frac{1}{i\hbar}[\hat{\mathbf{X}},\hat{H}].
\end{equation}
For the free particle, the~velocity is obtained, explicitly, as
\begin{equation}\label{Vfp}
\hat{\mathbf{v}}=\frac{1}{m}\left(\kappa^{\alpha}\hat{\mathbf{P}}+\kappa^{\beta}\hat{\mathbf{K}}_v\right),
\end{equation}
where
\begin{equation}\label{2Pr1}
\hat{\mathbf{K}}_v=\left(\hat{P}_Y-\hat{P}_Z\right)\mathbf{i}+\left(\hat{P}_X+\hat{P}_Z\right)\mathbf{j}
+\left(-\hat{P}_X+\hat{P}_Y\right)\mathbf{k}.
\end{equation}

It can be seen that there exists an intrinsic velocity, associated with the noncommutative parameters, and~driven by the stochastic fluctuations of the background geometry.
However, when the canonical momentum of the particle vanishes, $\left\langle\hat{\mathbf{P}}\right\rangle=0$, we obtain $\left\langle\hat{\mathbf{K}}_v\right\rangle=0$. 
In Section \ref{Sec.4.4} below, we give the Heisenberg representation of the velocity based on the SW map. This raises interesting issues regarding the nature of the principle of relativity
in noncommutative space, but~an in-depth discussion of these points lies outside the scope of the present review; for further literature, see references in Ref. \cite{Liang2}.

Consistent with the definition of velocity,
the acceleration is defined  as $\hat{\mathbf{a}}:=\frac{d}{dt}\hat{\mathbf{v}}$.
Again, by using the Heisenberg equation, this gives 
\begin{equation}\label{ap}
\hat{\mathbf{a}}=\frac{1}{i\hbar}[\hat{\mathbf{v}},\hat{H}].
\end{equation}
For the free particle, the~acceleration is obtained, explicitly, as
\begin{equation}\label{afp}
\hat{\mathbf{a}}=\frac{\eta}{m^2\hbar}(\kappa^\alpha+\kappa^\beta)\hat{\mathbf{K}}_p,
\end{equation}
where
\begin{equation}\label{KP2}
\hat{\mathbf{K}}_p=\left(\widehat{P}_y+\widehat{P}_z\right)\mathbf{i}-\left(\widehat{P}_x-\widehat{P}_z\right)\mathbf{j}
-\left(\widehat{P}_x+\widehat{P}_y\right)\mathbf{k}.
\end{equation}
Hence, there also exists an intrinsic stochastic acceleration.
We can interpret this intrinsic acceleration as arising from quantum fluctuations of the noncommutative phase space background.
This is associated with fluctuations in the components of the momentum, but~we note that the direction of the acceleration is not, in~general, the~same as the direction of the momentum~fluctuations.

\subsection{Seiberg--Witten Map and the Heisenberg~Representation} \label{Sec.4.4}

The noncommutative phase space provides a new operator algebra, beyond~the canonical Heisenberg algebra, with~which to explore unsolved puzzles in physics.
To establish a rigorous mathematical basis for the model, and~to allow clearer comparison with the canonical quantum formalism, several methods have been proposed in~the literature to~implement the noncommutative algebra (\ref{XYPP4a})-(\ref{XYPP4c}) via a map on the canonical quantum operators.
The three most used methods are the SW map, the~Moyal star product, and~the Wigner--Weyl phase space formulation~\cite{Bellucci,Gamboa,Gamboa1}.
Here, we outline the method of Seiberg and Witten (SW), which constructs a map linking the noncommutative phase space relations to the canonical Heisenberg algebra~\cite{Seiberg,Seiberg1,Bertolami1}.

Formally, the~SW map is defined as a map from the noncommutative phase space to the canonical Heisenberg phase space, which preserves the form of functions of the canonical variables, i.e.,
\begin{equation}\label{SWM1}
\phi_{\rm SW}:
\begin{array}{ccc}
\left(\hat{\mathbf{X}},\hat{\mathbf{P}}\right) & \rightarrow & \left(\hat{\mathbf{x}},\hat{\mathbf{p}}\right), \\
\hat{O}\left(\hat{\mathbf{X}},\hat{\mathbf{P}}\right) & \rightarrow & \hat{O}\left(\hat{\mathbf{x}},\hat{\mathbf{p}}\right),
\end{array}
\end{equation}
where $(\hat{\mathbf{X}},\hat{\mathbf{P}})$ obey the algebra (\ref{XYPP4a})-(\ref{XYPP4c}) and $(\hat{\mathbf{x}},\hat{\mathbf{p}})$ obey 
Equation (\ref{HAGA}).

{In other words, the~noncommutative relations between the operators in 
Equations (\ref{XYPP4a})-(\ref{XYPP4c}) can be implemented equivalently by~the Heisenberg commutation relations, together with the SW map.
The SW map may be implemented, explicitly, via the so-called Bopp shift~\cite{Bertolami1}, which is expressed as}
\begin{eqnarray}
\hat{X}^{i} = \hat{x}^{i} + \alpha^{ik}\hat{p}_{k}, \label{SWM2a}
\end{eqnarray}
\begin{eqnarray}
\hat{P}_{j} = \hat{p}_{j} + \beta_{jl}\hat{x}^{l}, \label{SWM2b}
\end{eqnarray}
where
\begin{eqnarray} \label{theta-eta}
\theta^{ij} := \hbar(\alpha^{ji} - \alpha^{ij}), \quad \eta_{ij} := \hbar(\beta_{ji} - \beta_{ij})
\end{eqnarray}
and
\begin{eqnarray} \label{kappa}
\kappa^{i}{}_{j} := \hbar(\delta^{i}{}_{j}  - \alpha^{ik}\beta_{jk}).
\end{eqnarray}

By using the matrices (\ref{ABC}), this gives
\begin{eqnarray} \label{NCXP2a}
\left(
\begin{array}{c}
\hat{X} \\
\hat{Y} \\
\hat{Z}
\end{array}
\right)&=& \left(
\begin{array}{c}
x -\frac{\theta}{2\hbar}\left(\hat{p}_{y}+\hat{p}_{z}\right) \\
y +\frac{\theta}{2\hbar}\left(\hat{p}_{x}-\hat{p}_{z}\right) \\
z +\frac{\theta}{2\hbar}\left(\hat{p}_{x}+\hat{p}_{y}\right)
\end{array}
\right), 
\end{eqnarray}
\begin{eqnarray} \label{NCXP2b}
\left(
\begin{array}{c}
\hat{P}_{X} \\
\hat{P}_{Y} \\
\hat{P}_{Z}
\end{array}
\right) &=& \left(
\begin{array}{c}
\hat{p}_{x} +\frac{\eta}{2\hbar}(y+z) \\
\hat{p}_{y} -\frac{\eta}{2\hbar}(x-z) \\
\hat{p}_{z} -\frac{\eta}{2\hbar}(x+y)
\end{array}%
\right),
\end{eqnarray}
which implies
\begin{equation}\label{LDa1}
\left[\alpha^{ij}\right]=\left(
\begin{array}{ccc}
0 &  -\frac{\theta}{2\hbar}& -\frac{\theta}{2\hbar} \\
\frac{\theta}{2\hbar} & 0 & -\frac{\theta}{2\hbar} \\
\frac{\theta}{2\hbar} & \frac{\theta}{3\hbar} & 0
\end{array}\right)
\end{equation}
and
\begin{equation}\label{PI1}
\left[\beta_{ij}\right]=\left(
\begin{array}{ccc}
0 &  \frac{\eta}{2\hbar}& \frac{\eta}{2\hbar} \\
-\frac{\eta}{2\hbar} & 0 & \frac{\eta}{2\hbar} \\
-\frac{\eta}{2\hbar} & -\frac{\eta}{2\hbar} & 0
\end{array}\right).
\end{equation}
This representation of the position and momentum operators can be regarded as the Heisenberg representation of noncommutative quantum mechanics.
The SW map provides an efficient way to modify the Heisenberg algebra, giving rise to a noncommutative phase space, even though the map is not unitary or canonical.

By using the SW map (\ref{SWM2a})-(\ref{SWM2b}) together with the noncommutative matrices (\ref{ABC}), we obtain the explicit form of the noncommutative parameters that generalises the position--momentum commutator as
\begin{eqnarray} \label{kappa_values}
\kappa^\alpha= \hbar\left(1+\frac{\theta\eta}{2\hbar^2}\right), \quad \kappa^\beta=\frac{\theta\eta}{4\hbar}, 
\end{eqnarray}
which implies that the noncommutative phase space can be described by two independent parameters, $\theta$ and $\eta$.

However, it should be remarked that there is actually no unique SW map between the noncommutative phase space and the canonical Heisenberg phase space.
The SW map with the Bopp shift, defined by Equations~(\ref{LDa1}) and (\ref{PI1}), provides a straightforward way to capture, approximately, the essential features of the noncommutative algebra in terms of the usual canonical Heisenberg algebra~\cite{Bertolami1}.

Before concluding this section, we again stress that the choice of independent parameters for the antisymmetric matrices, $\theta$ and $\eta$, and~for the symmetric matrix, $\kappa^{\alpha}$ and $\kappa^{\beta}$, is the simplest ``natural'' choice available.
Nonetheless, in~principle, the~SW map and noncommutative algebras do not require constant matrices.
Furthermore, for~the SW map with the Bopp shift, (\ref{LDa1}) and (\ref{PI1}), this 
relatively simple choice is truly ``minimal'', since even $\kappa^{\alpha}$ and $\kappa^{\beta}$ can be expressed in terms of $\theta$ and $\eta$, according to Equation~(\ref{kappa_values}).
In Section \ref{Sec.4.8} below, we associate the two noncommutative parameters, $\theta$ and $\eta$, with~the Planck length and the cosmological constant, respectively, which, it may be hoped, could provide a physical scenario for the emergence of dark energy from noncommutative geometry.

\subsection{The Schr\"{o}dinger Equation in Noncommutative Phase~Space} \label{Sec.4.5}

Based on the SW map, the~momentum operators in Equations (\ref{NCXP2a})-(\ref{NCXP2b}) can be rewritten as $\hat{\mathbf{P}}=\hat{\mathbf{p}}-\mathbf{A}^{\eta}$, where 
\begin{subequations}\label{EANC1}
\begin{eqnarray}
A^{\eta}_{X} &=& -\frac{\eta}{2\hbar} (y+z), \\
A^{\eta}_{Y} &=& \frac{\eta}{2\hbar} (x-z),\\
A^{\eta}_{Z} &=& \frac{\eta}{2\hbar} (x+y),
\end{eqnarray}
\end{subequations}
and the $A^{\eta}_i$ can be regarded as the components of an effective gauge potential.
Thus, the~Schr\"{o}dinger equation for the free particle can be expressed as
\begin{equation}\label{SESW1}
i\hbar\frac{\partial \psi}{\partial t}=\frac{1}{2m}\left(\hat{\mathbf{p}}-\mathbf{A}^{\eta}\right)^2\psi.
\end{equation}

It can be seen that the Schr\"{o}dinger equation for a free particle in noncommutative phase space is analogous to that obtained, in~the canonical theory, for~a charged particle in an electromagnetic potential.
The continuity equation is given by
\begin{equation}\label{CE1}
\frac{\partial \rho}{\partial t}+\nabla\cdot \mathbf{J}=0,
\end{equation}
where $\rho=|\psi |^2$ is the probability density and the probability current density is defined as $\mathbf{J}=\frac{1}{m}\left\langle \psi\left|\hat{\mathbf{p}}-\mathbf{A}^{\eta}\right|\psi\right\rangle$.
It should be noted that the current density is not equal to the velocity given by Equation~(\ref{Vfp}).

\subsection{Noncommutative Gauge Fields and the Pauli~Equation} \label{Sec.4.6}

We suppose that spin in noncommutative phase space has the same form as in canonical quantum mechanics, because~spin is an intrinsic property of the particle, which is independent of the spacetime background~\cite{Armin}.
Thus, the~Pauli equation for a particle with nonzero spin and charge $q$, in~the noncommutative space, may be written as
\begin{equation}\label{PlE1}
i\hbar\frac{\partial \psi}{\partial t}=\left[\frac{1}{2m}\left[\boldsymbol{\sigma}\cdot\left(\widehat{\mathbf{p}}-\mathbf{A}^{\rm nc}\right)\right]^{2}+q\phi 
\right]\psi,
\end{equation}
where $\boldsymbol{\sigma} = \sigma_x {\bf i} + \sigma_y {\bf j} + \sigma_z {\bf k}$ and $\phi$ is the electric potential. 
The effective vector potential contains two terms,
\begin{equation}\label{ANC}
\mathbf{A}^{\rm nc}=\mathbf{A}+\mathbf{A}^{\eta},
\end{equation}
where $\mathbf{A}$ is the canonical vector potential and $\mathbf{A}^{\eta}$ is the effective contribution arising from the noncommutativity of the momentum components.
The total effective magnetic field is $\mathbf{B}^{\rm nc}=\nabla\times\mathbf{A}^{\rm nc}$, where 
$\mathbf{B}^{\rm nc}=\mathbf{B}+\mathbf{B}^{\eta}$ 
with
$\mathbf{B}=\nabla\times\mathbf{A}$ and $\mathbf{B}^{\eta}=\nabla\times\mathbf{A}^{\eta}$.
Thus, the~Pauli equation can be rewritten as
\begin{equation}\label{PlE2}
i\hbar\frac{\partial \psi}{\partial t}=\left[\frac{1}{2m}\left(\widehat{\mathbf{p}}-\mathbf{A}^{\rm nc}\right)^{2} 
-\frac{\hbar q}{2m}\boldsymbol{\sigma}\cdot \mathbf{B}^{\rm nc}
+q\phi \right]\psi.
\end{equation}

By analogy with canonical electrodynamics, we also define the effective electromagnetic field tensor, generated from the effective 
gauge potential in the noncommutative phase space, as~\begin{equation}\label{GF1}
F_{\mu\nu}^{\eta}=\partial_{\mu} A^{\eta}_{\nu} - \partial_{\nu} A^{\eta}_{\mu}.
\end{equation}
Substituting the gauge potential (\ref{EANC1}) into 
Equation (\ref{GF1}), the~explicit form of $F_{\mu\nu}^{\eta}$ is obtained as 
\begin{equation}\label{GF2}
\left[F^\eta_{\mu\nu}\right]=\left(
\begin{array}{ccc}
 0 & 
-{\eta}/{\hbar} & -{\eta}/{\hbar}\\
{\eta}/{\hbar} & 0 & -{\eta}/{\hbar}\\
{\eta}/{\hbar} & {\eta}/{\hbar} & 0
\end{array}
\right).
\end{equation}
Interestingly, the effective gauge field is induced by the noncommutativity between different directional components of momentum, in the noncommutative phase space.
This noncommutative effect can be interpreted as a spacetime curvature, so that the resulting gauge field can be naturally related to the value of the scalar curvature.
For this reason, we relate $\eta$ to the observed value of the cosmological constant, in~Section~\ref{Sec.4.8}.

Following Equation (\ref{SWM1}), any observable operator in the noncommutative phase space can be mapped to the Heisenberg representation, based on the SW map. 
Actually, the wave function in noncommutative phase space is also given by $\Psi\left(\mathbf{X}\right)=\left\langle \mathbf{X}|\Psi\right\rangle$ or $\tilde{\Psi}\left(\mathbf{P}\right)=\left\langle \mathbf{P}|\Psi\right\rangle$, by~analogy with the canonical theory.
However, as~a first-order approximation, it is convenient to assume 
that~\cite{Harko,Liang,Liang1}
\begin{eqnarray}  \label{SW_psi-A}
\Psi\left(\mathbf{X}\right) \simeq \psi(\mathbf{x})+\mathcal{O}(\theta), \label{SW_psi-x}
\end{eqnarray}
\begin{eqnarray}  \label{SW_psi-B}
\Psi\left(\mathbf{P}\right) \simeq \psi(\mathbf{p})+\mathcal{O}(\eta). \label{SW_psi-p}
\end{eqnarray}

\subsection{Noncommutativity-Induced Anomalous Velocity and~Acceleration} \label{Sec.4.7}

For the free particle, using the SW map in Equations (\ref{SWM1}) and (\ref{SWM2a})-(\ref{SWM2b}), the~{expectation value of the velocity} in the Heisenberg representation can be expressed as
\begin{equation}\label{Vfp2}
\langle\hat{v}^i\rangle=\frac{1}{\hbar m}\left(\kappa^i_j\langle\hat{p}_j\rangle+\frac{\eta}{2}\gamma^i_j\langle x^j\rangle\right), 
\end{equation}
where
\begin{equation}\label{PPR3}
\left[\gamma^i_j\right]=\left(
\begin{array}{ccc}
0 & \left(1+\frac{3\theta\eta}{4\hbar^2}\right) & \left(1+\frac{3\theta\eta}{4\hbar^2}\right)\\
-\left(1+\frac{3\theta\eta}{4\hbar^2}\right) & 0 & \left(1+\frac{3\theta\eta}{4\hbar^2}\right) \\
-\left(1+\frac{3\theta\eta}{4\hbar^2}\right) & -\left(1+\frac{3\theta\eta}{4\hbar^2}\right) & 0 
\end{array}
\right).
\end{equation}
Similarly, the~{expectation value of the acceleration} is obtained as
\begin{equation}\label{force2}
\langle\hat{a}_i\rangle=\frac{1}{m^2\hbar}\left(1+\frac{3\theta\eta}{4\hbar^2}\right)
\left(\eta_{ij}\langle \hat{p}_{j}\rangle-\frac{\eta^2}{\hbar}\chi_{ij}\langle x^j\rangle\right),
\end{equation}
where 
\begin{equation}\label{PP3}
\left[\chi_{ij}\right] =\left(
\begin{array}{ccc}
1 & 
{1}/{2} & -{1}/{2}\\
{1}/{2} & 1 & {1}/{2}\\
-{1}/{2} & {1}/{2} & 1\\
\end{array}\right).
\end{equation}

As noted previously, in~Section~\ref{Sec.4.3}, the~{velocity and acceleration} contain an intrinsic stochastic perturbation arising from the noncommutativity of momenta in different directions of the phase space.
Here, these are expressed in terms of the two independent parameters {of our specific implementation of the SW map, $\theta$ and $\eta$.
This anomalous acceleration can be interpreted as a quantum effect, since it is induced by the noncommutative algebra. 
In Section \ref{Sec.4.8} below, we express the noncommutative parameters in terms of the Planck length and the cosmological constant, linking the noncommutative algebra with the spacetime background and nonzero minimal energy~density.

\subsection{Physical Interpretations of the Noncommutative~Parameters} \label{Sec.4.8}

In canonical quantum mechanics, Planck's constant plays an essential role in quantizing the phase space of elementary particles, such that quantum states are described by state vectors in a Hilbert space.
Roughly speaking, $\hbar$ represents the minimum (incompressible) volume of a phase space fluid element, $\Delta X^{i}\Delta P_{j}$ \cite{Carlo,Thomas}.
In noncommutative phase space, the~noncommutative parameters $\theta$ and $\eta$ play analogous roles with respect to physical space and momentum space, respectively, inducing minimum bounds on the volumes $\Delta X^{i}\Delta X^{j}$ and $\Delta P_{i}\Delta P_{j}$, which cannot be further compressed below their minimum~values.

Actually, one can adopt different parametrization schemes for the noncommutative parameters, for~various physical problems, associated with different energy and spacetime scales.
In other words, how one endows the noncommutative parameters with physical meanings (and scales) depends on what physical problems are being addressed. 
Therefore, let us note that many studies suggest that spacetime should be quantized, inducing a minimum length scale of the order of the Planck length~\cite{Fredenhagen}.
Similarly, the~emergence of dark energy implies that there exists a minimum curvature of spacetime, expressed in terms of the cosmological constant~\cite{Sabino,Peebles}, which can be naturally associated with the nonzero energy/momentum density obtained in a noncommutative phase space model.
(We recall that a minimum positive curvature is equivalent, according to the gravitational field equations, to~a minimum positive energy density.)}

In particular, gedanken experiments in quantum gravity, together with various theoretical approaches, such as string theory~\cite{Fredenhagen}, loop quantum gravity~\cite{Carlo,Thomas}, and~others, suggest spacetime quantization at the Planck scale.
While the status of a minimum possible momentum is less clear, several models also propose this~\cite{Sabino,Peebles}, and~it is worth noting that in~a universe governed by dark energy there exists a finite de Sitter horizon~\cite{Sabino,Peebles}.
This places on upper bound on the value of a particle's de Broglie wavelength and, hence, a~lower value on its momentum uncertainty, of~the order of the de Sitter momentum, $m_{\rm dS}c = \hbar\sqrt{\Lambda/3}$, mentioned previously in Section~\ref{Sec.2.3}.

Thus, we propose a parametrization scheme for the noncommutative parameters, which is associated with the Planck length and the cosmological constant, namely
\begin{equation}\label{PMTT}
\theta=\ell_{\rm Pl}^2,  \quad \eta= m_{\rm dS}^2c^2,
\end{equation}
where we define the Planck and de Sitter mass scales explicitly as~\begin{eqnarray}\label{Planck-deSitter}
\ell_{\rm Pl} = \sqrt{\frac{\hbar G}{c^3}} \simeq 10^{-33} \, {\rm cm}, \quad m_{\rm Pl} = \sqrt{\frac{\hbar c}{G}} \simeq 10^{-5} \, {\rm g}, \\ \label{Planck-deSitter-a}
\ell_{\rm dS} = \sqrt{\frac{3}{\Lambda}} \simeq 10^{28} \, {\rm cm}, \quad m_{\rm dS} = \frac{\hbar}{c}\sqrt{\frac{\Lambda}{3}} \simeq 10^{-66} \, {\rm g}. \label{Planck-deSitter-a} 
\end{eqnarray}

Combining the relations (\ref{PMTT}) and (\ref{Planck-deSitter}), we obtain the independent components of the generalised 
position--momentum commutator, in~the noncommutative phase space, as
\begin{eqnarray} \label{kappa_Planck-deSitter}
\kappa^\alpha=\hbar\left(1 + \Delta\right), \quad \kappa^\beta = \frac{\hbar\Delta}{2}, 
\end{eqnarray}
where
\begin{eqnarray} \label{delta}
\Delta := \frac{\rho_{\Lambda}}{\rho_{\rm Pl}} \simeq 10^{-122},
\end{eqnarray}
and
\begin{eqnarray} \label{beta}
\rho_{\Lambda} = \frac{\Lambda c^2}{8\pi G}, \quad \rho_{\rm Pl} = \frac{3}{4\pi}\frac{c^5}{\hbar G^2},
\end{eqnarray}
are the dark energy and Planck densities, respectively, with $G$ the gravitational constant. 
Interestingly, a~new model of generalised uncertainty relations was developed~\cite{Lake1,Lake2}, in~which the action scale $\beta := 2\hbar\sqrt{\Delta} \simeq \hbar \times \mathcal{O}(10^{-61})$ was proposed as the fundamental quantum of action for spacetime, as~opposed to matter.
In this model, the~generalised uncertainties were obtained without modifying the canonical space--space or momentum--momentum commutators, but the parameterization above suggests that an extension of this approach could be used to provide a physical mechanism for noncommutative geometry.
This intriguing possibility will be analysed in~detail in~a future~work and we note that similar ideas, in which the noncommutativity parameters for position and mometum space are identified with the Planck and de Sitter scales, respectively, were also presented in a different model based on Snyder-de Sitter space \cite{Franchino-Vinas:2021bcl,Franchino-Vinas:2019nqy}.

\section{Conclusions and~Outlook}  \label{Sec.5}

The interplay between physics and mathematics stimulates new ideas to resolve unsolved puzzles in nature.
In this review, we have given a brief introduction to the interplay between physics and mathematics in the fields of noncommutative geometry and noncommutative phase space, encompassing topics in both classical physics and quantum~mechanics.

Although noncommutative phenomena were discovered, even in classical mechanics, their true significance became apparent only with the advent of quantum theory in the 1920s. Since then, noncommutative structures have inspired bold new attempts to solve unsolved problems in gravity and high-energy theory, and, arguably, their significance to modern theoretical physics has only increased since the heady days of the early twentieth~century.

From classical Poisson brackets to the Heisenberg commutation relations and the quantum Hall effect, physicists have found real-world phenomena described by noncommutative algebras and geometries.
Furthermore, as~we attempt to extend our current physical theories into previously unprobed regimes at the Planck scale, or~dark energy scale, infinities and singularities unavoidably emerge, suggesting new types of noncommutative structures which may be able to cure them. Increasingly, a~large number of researchers believe that the infinities and singularities which break all known physical laws may be cured if there exist quantum fluctuations of the spacetime background, governed by noncommutative algebras, which are able to prevent such divergences.
Thus, noncommutative phenomena inspire many attempts to construct a unified framework for both gravity and high-energy particle physics~\cite{Douglas,Szabo,Konechny}.

In this review, we have introduced the basic concepts underlying noncommutative phenomena in classical and quantum mechanics and~presented some important examples in condensed matter and statistical physics.
We discussed the basic noncommutative algebras that arise in physical theories, including the classical Poisson brackets in symplectic geometry, the~Heisenberg algebra of fundamental operators, acting on the Hilbert space of canonical quantum mechanics, and~the Lie, Clifford, and Dirac algebras associated with rotational symmetry and spin.
On a more theoretical note, we also gave brief expositions of the Snyder and Nambu algebras, which have been proposed as extensions of existing physical theories, and~are intended to help cure the emergence of the singularities mentioned above~\cite{Sabino,Peebles}.

Based on the SW map, we outlined the basic properties and novel phenomena that occur in the noncommutative extension of the Heisenberg phase space, incorporating both space--space and momentum--momentum noncommutativity.
These include the breaking of translation and rotational symmetries, as~well as important phenomenological predictions like the existence of anomalous, stochastic perturbations to the velocity and acceleration of free particles, induced by noncommutativity.
The stochastic perturbations can be viewed as an additional quantum force, driving particle motion due to quantum fluctuations of the background geometry, and~we showed that the noncommutative terms give rise to an effective gauge field in the Schr\"{o}dinger and Pauli equations. With~this in mind, we proposed a parametrization scheme for the noncommutative parameters, which associated them with both the Planck length and the dark energy density, where the latter is expressed in terms of the cosmological~constant.

Based on this parametrization scheme, the~effective gauge field that arises from the noncommutativity of the phase space can also be interpreted in terms of the minimum length, and~minimum energy density of the universe.
We showed that this gives rise to phenomenologically interesting effects on the dynamics of free particles, which are subjected to intrinsic stochastic velocity and acceleration perturbations.
These perturbations depend on the initial momentum and position, and~can be regarded as a quantum effects induced by the noncommutative phase space.
The quantum anomalous acceleration of free particles could actually provide a microphysical model for dark~energy.

However, we note that noncommutative models in physics have been developed in many different ways, including Conne's approach to noncommutative 
geometry~\cite{Connes,Connes1}, 
noncommutative M-theory~\cite{Konechny,Emil}, noncommutative field theory based on the Moyal star product~\cite{Szabo}, the~principle of relative locality, and~$\kappa-$deformed spacetime symmetries based on Hopf algebras~\cite{Giovanni,Alessandra,Michele}.
These interesting results suggest that noncommutative phase space may provide a deep connection between the dynamics of microscopic particles and the quantum theory of gravity, including dark 
energy; for details of these and~other approaches to noncommutative phenomena, that were not covered in the present  pedagogical introduction, see the bibliography and references~therein.

\section*{Acknowledgments}

M.L. thanks the Department of Physics and Materials Science, Faculty of Science, Chiang Mai University, for providing research facilities. 
This work was supported by the Natural Science Foundation of Guangdong Province (P.R. China), grant no. 2021A1515010036. 



\end{document}